\documentclass{aa}  

\usepackage{graphicx}
\usepackage{txfonts}
\usepackage{multirow}
\usepackage{amsmath}
\usepackage{comment}
\usepackage{csquotes}
\usepackage{siunitx}

\usepackage{xcolor}

\DeclareSIUnit \solarmass {M_{\odot}}
\DeclareSIUnit \parsec {pc}
\DeclareSIUnit \erg {erg}

\usepackage{hyperref}
\begin{document}

   \title{Prospects for detecting generic fast-time features in the neutrino lightcurve of nearby supernovae in neutrino telescopes}

   \titlerunning{Detecting fast-time features in supernova neutrinos}

   \subtitle{}

   \author{J. Beise\inst{1}
          \and
          S. BenZvi\inst{2}
          \and
          S. Griswold\inst{3}
          \and
          N. Valtonen-Mattila\inst{4}
          \and
          E. O'Sullivan\inst{5}
          }

   \institute{Dept. of Physics and Astronomy, Uppsala University, Box 516, SE-75120                Uppsala, Sweden\\ \email{jakob.beise@physics.uu.se}\label{inst1}
              \and
              Dept. of Physics and Astronomy, University of Rochester, Rochester, NY 14627, USA\\ \email{sbenzvi@ur.rochester.edu}\label{inst2}
              \and
              Dept. of Physics and Astronomy, University of Rochester, Rochester, NY 14627, USA\\ \email{sgriswol@ur.rochester.edu}\label{inst3}
              \and
              Fakultät für Physik \& Astronomie, Ruhr-Universität Bochum, D-44780 Bochum, Germany\\ \email{nvalto@astro.ruhr-uni-bochum.de}\label{inst4}
              \and
              Dept. of Physics and Astronomy, Uppsala University, Box 516, SE-75120 Uppsala, Sweden\\ \email{erin.osullivan@physics.uu.se}\label{inst5}
         }

   %\date{Received September 15, 1996; accepted March 16, 1997}

  \abstract{
  Neutrino emission offers a direct probe into the hydrodynamics and energy transport processes within a supernova. Fast-time variations in the neutrino luminosity and mean energy can provide insights into phenomena like turbulence, convection, and shock revival.
  In this paper, we explore the detection capabilities of large-volume neutrino telescopes such as the IceCube Neutrino Observatory and the planned IceCube-Gen2 detector in identifying generic fast-time features in the neutrino light curve. We also investigate the potential enhancement in detection sensitivity using wavelength shifters, which can improve light collection efficiency. 
  By employing a Short-Time Fourier Transform analysis, we quantify the excess power in the frequency spectrum arising from fast-time modulations and compute the detection horizon for a range of generic models.
  We find that with IceCube we can already see the strongest modulation models (>50\% amplitude) for progenitors located anywhere in the Milky Way. Sensitivity to weaker modulations (>20\% amplitude) is possible in future detectors like IceCube-Gen2, in particular with the use of wavelength shiftersFor all detector configurations, the frequency and central time of the fast-time feature at the 5$\sigma$ detection horizon can be measured with a resolution of \SI{7.0}{\Hz} and \SI{17}{\milli \s} respectively.
  }

   \keywords{core collapse supernova --
                neutrinos --
                fast-time features --
                telescope
               }

   \maketitle
%
%________________________________________________________________

\section{Introduction}
\label{sec:intr}

The explosion of core-collapse supernovae (CCSNe) releases over 99\% of the gravitational binding energy of the stellar core in the form of $\mathcal{O}$(\SI{10}{\mega \eV}) neutrinos. CCSNe are among the few confirmed astrophysical sources of neutrinos, as evidenced by the direct detection of neutrinos from SN1987A \citep{SN1987A_Kamiokande, SN1987A_IMB, SN1987A_Baksan}. The neutrinos not only carry unique information about the internal conditions in the core, but also appear to play a critical role in the transfer of energy to infalling material, driving the collapsing star to a supernova explosion \citep{Bethe:1990heat}. However, while the neutrino-driven explosion mechanism is widely accepted, the internal dynamics of CCSNe remain an open question. Multi-dimensional simulations have demonstrated that certain phenomena, such as convective overturn in the neutrino-heated layer \citep{Herant:1992conv, Herant:1994conv, Burrows:1995conv, Janka:1995conv, Janka:1996conv}, the standing-accretion shock instability (SASI) \citep{Blondin:2003sasi, Blondin:2006sasi, Foglizzo:2006sasi, Foglizzo:2007sasi, Laming:2007sasi}, and rapid rotations \citep{Janka:2016snro, Takiwaki:2016snro, Summa:2018snro}, can drive fast-time modulations in the neutrino emission during the accretion phase and aid the explosion mechanism. These periodic modulations, which occur on timescales of $\mathcal{O}$(\SI{100}{\milli \s}) and periodicities of $\mathcal{O}$(\SI{10}{\milli \s}), are of particular interest because they encode valuable information about the internal structure and processes within the supernova core, including rotation speed, shock instabilities, convection, neutrino flavour conversion, and energy deposition of neutrino transport mechanisms. Since the rate of galactic CCSNe is estimated to be a few per century \citep{Rozwadowska:2021rate, Adams:2013rate} it is  to capture as much information as possible from this rare event. 

For this analysis, we consider long-string water Cherenkov detectors (described in Sec.~\ref{sec:ana_ic}) as they have the capability to measure fast time features with high statistics. The IceCube Neutrino Observatory \citep{IceCube:2016inst} can observe a CCSN in our galaxy above $10\sigma$ for all progenitor masses and has sensitivity to heavier progenitors in the Large and Small Magellanic Cloud (LMC and SMC) (\SI{60}{\kilo\parsec}) \citep{IceCube:2011pros, IceCube:2024ccsn}. KM3NeT's ARCA and ORCA detectors \citep{KM3NeT:2016inst} have $>3\sigma$ sensitivity for all progenitor masses, with higher sensitivity to heavier progenitors, extending to the LMC \citep{KM3NeT:2021ccsn}.

Fast-time features such as the Standing Accretion Shock Instability (SASI), rotation, and convection-driven modulations in CCSNe have been previously modelled in various detectors. Detection horizons for SASI modulations in IceCube vary, with estimates ranging from \SI{8}{\kilo\parsec} to \SI{35}{\kilo\parsec}, depending on factors such as signal amplitude, progenitor mass, and search methods \citep{Tamborra:2014sasi, Walk:2020sasi, Beise:2023sasi}. Next-generation neutrino detectors such as IceCube-Gen2 (described further in Sec.~\ref{sec:ana_ic}) improve sensitivity to these modulations, particularly for detecting fainter signals. Rotating progenitors can produce an imprint on both gravitational waves and neutrinos with distinct frequency modulations that are detectable by IceCube from \SI{1}{\kilo\parsec} to \SI{5}{\kilo\parsec} \citep{Walk:2018snft, Westernacher-Schneider:2019snro}, though the sensitivity is model dependent. Other detectors like KM3NeT and Hyper-Kamiokande \citep{Abe:2011hypk} can also detect fast-time features. KM3NeT has a detection horizon of \SI{3}{\kilo\parsec} to \SI{8}{\kilo\parsec} for SASI signals \citep{KM3NeT:2021ccsn, Molla:2019snft}, and modulations of the neutrino lightcurve would be visible for Hyper-Kamiokande up to the Galactic Centre \citep{Migenda:2015snft, Yano:2021hypk}.

This paper aims to investigate the detection of generic periodic fast-time features imprinted on a given model of the neutrino lightcurve. We consider the sensitivity of the IceCube detector, as well as IceCube-Gen2, a proposed $10~\mathrm{km}^3$ extension of the IceCube Neutrino Observatory. We consider both a baseline design of IceCube-Gen2 as well as a design that incorporates wavelength-shifting sensors (see Sec.~\ref{sec:ana_ic}). In Sec.~\ref{sec:ana} we describe the methodology and analysis. We present and discuss our results in Sec.~\ref{sec:res} and Sec.~\ref{sec:disc}, respectively, and summarise our findings in Sec.~\ref{sec:conc}.

%__________________________________________________________________

\section{Analysis}
\label{sec:ana}

%This section covers the general detection technique of long-strong water Cherenkov detectors, in particular the IceCube detector (Sec.~\ref{sec:ana_ic}), and describes supernova detection (Sec.~\ref{sec:ana_det}) as well as simulation details (Sec.~\ref{sec:ana_sim}). In Sec. ~\ref{sec:ana_hypo} we define the hypothesis test while we explain the Fourier analysis in Sec.~\ref{sec:ana_four}. We describe our statistical method in Sec.~\ref{sec:ana_ts} and explain how to obtain the detection significance in Sec.~\ref{sec:ana_sig}. Finally, we detail how to reconstruct model parameters from the Fourier transform in Sec.~\ref{sec:ana_reco}.

\subsection{Long-string water Cherenkov detectors and IceCube}
\label{sec:ana_ic}

Long-string Cherenkov detectors such as IceCube \citep{IceCube:2016inst} and KM3NeT \citep{KM3NeT:2016inst} sparsely instrument $\mathcal{O}(1~\SI{}{km}^3)$ volumes of ice or water with optical sensors equipped with photomultiplier tubes (PMTs). The sensors are attached to cables that provide structural support to the optical modules and supply power to each PMT. At MeV energies, these neutrino telescopes are primarily sensitive to the $\overline{\nu}_e$ flux from CCSN through the production of positrons in the inverse-beta decay (IBD) on protons in water (or ice) molecules. While positrons produced in IBD travel $\sim0.5~\SI{}{cm}(E_{e^+}/\SI{}{MeV})$ \citep{IceCube:2011pros}, the sensor spacing in the detector is typically tens of meters. As a result, the Cherenkov photons produced by a given IBD positron are typically detectable in at most one optical sensor. In ice, the Cherenkov light is observed on top of sensor backgrounds, which include each PMT's dark current, radioactive decays in the PMT and sensor glass, and Cherenkov light produced by coincidence atmospheric muons. In seawater detectors such as KM3NeT, bioluminescence is an additional source of background. As a result, km$^3$ Cherenkov detectors cannot reconstruct individual CCSN neutrinos, but instead observe the signal as an excess of the collective event rate above the background during the $\mathcal{O}$(\SI{10}{\s}) accretion and cooling phase of the CCSN. Due to the high signal statistics and $\mathcal{O}(1~\SI{}{ns})$ time resolution of long-string water Cherenkov detectors, they can measure fast-time features in the neutrino lightcurve with high precision. Sensor designs featuring multi-PMT sensors can reduce the background rate by requiring module coincidences at the expense of some signal.

IceCube \citep{IceCube:2016inst} instruments \SI{1}{\kilo \m^3} of glacial ice at the geographical South Pole and consists of 5160 optical sensors on 86 vertical cables, or \enquote{strings}, buried \SI{1450}{\m} to \SI{2450}{\m} underneath the ice (see Fig.~\ref{fig:geom}, right). The central 7 strings are deployed in a dense configuration known as DeepCore \citep{IceCube:2011ucd}. The IceCube Upgrade \citep{Ishihara:2019}, currently being deployed, extends the dense infill array with roughly 700 new sensors in the DeepCore region. The Upgrade is optimised to improve the detector calibration and carry out precise atmospheric neutrino oscillation measurements. IceCube-Gen2 \citep{IceCube-Gen2:2023tdr} is an envisioned large-scale extension supplementing the current geometry by a total of 9,600 new optical sensors on 120 new \enquote{strings} with 80 modules each, spanning depths between \SI{1369}{\m} and \SI{2689}{\m} (see Fig.~\ref{fig:geom}, left). The sensors in IceCube-Gen2 will be segmented, housing multiple small-diameter PMTs. Given the increased photocathode area, the IceCube-Gen2 design will enhance photon collection and suppress sensor backgrounds. This design will extend the sensitivity of the observatory to low-mass progenitor CCSNe out to the LMC and SMC \citep{LozanoMariscal:2021mult, IceCube-Gen2:2023tdr}. 

Wavelength shifters (WLS) are low-noise sensors that can augment photon collection through increased photon collection area and UV sensitivity and are therefore a promising technology for low signal-to-noise ratio measurements such as supernova neutrino detection. The Wavelength-Shifting Optical Module (WOM) \citep{WOM:2022ini, WOM:2022upgr} is a demonstrator technology being deployed with several sensors in the IceCube Upgrade. WLS tubes, operating without electronics or PMT readout, are a potential add-on to the already-planned segmented sensors in IceCube-Gen2 (see Fig.~\ref{fig:geom}, left) as a cost-efficient, low-noise, and scalable photon collector.

\subsection{Supernova detection in IceCube}
\label{sec:ana_det}

We determine the time-dependent number of detected signal photons in a time interval $\Delta t$ by summing over all sensor types $i$, where each term is given by the number of sensors of type $i$, $m_i$, multiplied by the average per-sensor hit rate $\langle R_{\text{SN}, i} (t)\rangle$, integrated over the interval:
\begin{equation}
     N_{\text{SN}} = \sum_i m_i \int_{0}^{\Delta t} dt \ \langle R_{\text{SN}, i} (t) \rangle \ .
\end{equation}
The average single-sensor hit rate for a specific reaction and target is defined as
\begin{equation}
\begin{split}
    \langle R_{\text{SN}, i}(t)\rangle &=  \langle \epsilon_{\tau, i} \rangle \times \frac{n_{\text{target}} \  \mathcal{L}_{\text{SN}}^{\nu}(t)}{4\pi d_{\text{SN}}^2 \langle E_{\nu} (t) \rangle} \int_{0}^{\infty} dE_{e} \int_{0}^{\infty} dE_{\nu} \\
    & \times \frac{d \sigma}{d E_{e}} (E_{e},\ E_{\nu}) \ N_{\gamma}(E_{e}) \ V_{\gamma, i}^{\text{eff}} \ f(E_{\nu}, \langle E_{\nu} \rangle, \alpha_{\nu}, t) \ ,
\end{split}
\end{equation}
where $\langle \epsilon_{\tau, i} \rangle$ is the sensor-averaged dead time efficiency, $n_{\text{target}}$ is the density of targets in ice, $d_{\text{SN}}$ is the distance to the supernova with neutrino luminosity $\mathcal{L}_{\text{SN}}^{\nu}(t)$ and normalised neutrino energy distribution $f(E_{\nu}, \langle E_{\nu} \rangle, \alpha_{\nu}, t)$ depending on the mean energy $\langle E_{\nu} \rangle$ and the shape parameter $\alpha_{\nu}$. $V_{\gamma, i}^{\text{eff}}$ is the effective volume of sensor $i$ for a single photon, defined as the hypothetical volume of the sensor with 100\% detection efficiency, $N_{\gamma}(E_{e})\approx 178 \cdot E_{e}$ is the energy-dependent number of radiated Cherenkov photons and $\frac{d \sigma}{d E_{e}} (E_{e},\ E_{\nu})$ is the differential cross section for producing a positron or electron of energy $E_e$ from a neutrino with energy $E_{\nu}$. For further reading, we refer to Refs.~\citep{IceCube:2011pros, IceCube:2024ccsn}.

The dead time efficiency is sensor-dependent and a function of the detector hit rate and the non-paralysing dead time $\tau$: 
\begin{equation}
    \epsilon_{\tau, i}(R_{\text{SN}}) = \frac{\epsilon_{\tau, i}^{\text{max}}}{(1+R_{\text{SN}} \cdot \tau)} \ ,
\end{equation}
where $\epsilon_{\tau, i}^{\text{max}}$ is the maximum dead time efficiency obtained from simulation. The number of background hits is the sum over all sensor types $i$, where each term is the product of $m_i$, the time window $\Delta t$ and the averaged per-sensor noise rate $\langle R_{\tau, i} \rangle$:
\begin{equation}
    N_{\text{bkg}} = \sum_i m_i \ \Delta t \  \langle R_{\tau, i} \rangle \ .
\end{equation}
The dead time $\tau$ = \SI{250}{\micro \s} suppresses the noise rate $\langle R_{\tau, i} \rangle$ while retaining the majority of the signal.

\subsection{Detector simulation}
\label{sec:ana_sim}

\begin{figure*}
    \centering
    \includegraphics[width=0.7\linewidth]{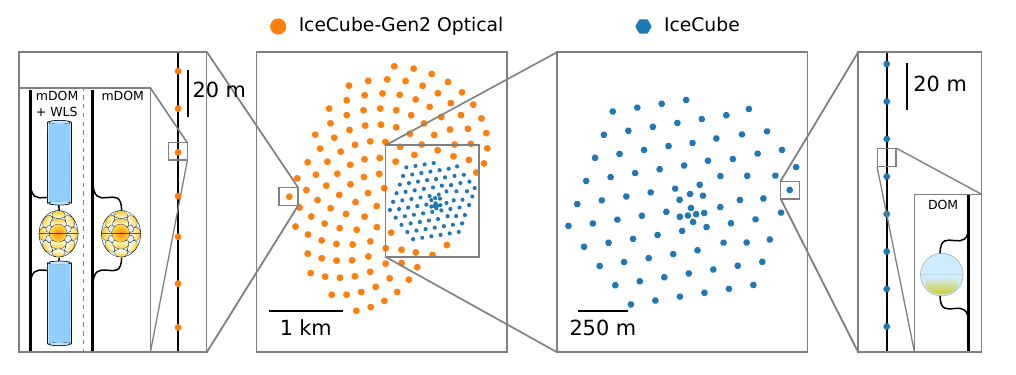}
    \caption{Detector geometry used in this study. \textit{Middle right}: Top view of the current IceCube detector layout. \textit{Right}: Sketch of the DOM sensor. \textit{Middle left}: Top view of the planned, sunflower-shaped IceCube-Gen2 optical array. \textit{Left}: Sketch of the mDOM with and without wavelength-shifting tube. Middle panel adapted from \citet{IceCube-Gen2:2023tdr}.}
    \label{fig:geom}
\end{figure*}

In this paper, we consider three detector scenarios (see Fig.~\ref{fig:geom}) which we will refer to as IceCube (including DeepCore), Gen2 (IceCube-Gen2 excluding WLS) and Gen2+WLS (IceCube-Gen2 including WLS). To estimate the sensitivity of IceCube-Gen2, we assume a design using the multi-PMT Digital Optical Module (mDOM) \citep{Kossatz:2017mdom, IceCube-Gen2:2023tdr}; however, alternative Gen2 sensors are currently under development. The WLS component of this study considers a \SI{2}{\m} long tube with an outer diameter of \SI{256}{\milli \m}, a tube thickness of \SI{10}{\milli \m} and the same material properties as the inner tube of the WOM \citep{WOM:2022ini}. 

\begin{table}
    \caption{\label{tab:char}Characteristics of the sensors considered in this study.}
    \centering
    \begin{tabular}{lcccc}
    \hline\hline
    Geometry & $m$ & $\langle V_{\gamma}^{\text{eff}} \rangle$ [\SI{}{\m^{3}}] & $\langle R_{\tau} \rangle$ [Hz] & $\epsilon_{\tau}^{\text{max}}$ [\%]\\
    \hline
    \multirow{2}{*}{IceCube} & 4800 &  0.17 & 285 & 88.3\\
    & 360 & 0.23 & 359 & 84.6\\
    \hline
    Gen2 & 9760 & 0.33 & 2300 & 95.8\\
    Gen2+WLS & 9760 & 0.60 & 2700 & 95.8\\
    \hline
    \end{tabular}
    \tablefoot{For Gen2 and Gen2+WLS we consider the IceCube geometry together with the indicated number of additional sensors. $m$ is the number of sensors. $\epsilon_{\tau}^{\text{max}}$ is simulated for a \SI{250}{\micro \s} dead time.}
\end{table}

Table~\ref{tab:char} lists all relevant data for the modules considered in this study. For IceCube, we simulate 4800 DOMs and 360 high quantum efficiency (HQE) DOMs. For Gen2, we add 9760 additional mDOMs on 122 \enquote{strings}. This selection differs slightly from the baseline design of 120 \enquote{strings} described in \citet{IceCube-Gen2:2023tdr}, but the choice is not expected to impact the results of this paper. In the case of Gen2+WLS, the mDOMs are complemented by two WLS tubes: one above and one below the mDOM (see Fig.~\ref{fig:geom}). The WLS component nearly doubles the photon effective volume while only modestly increasing the noise rate by \SI{400}{\Hz}, equivalent to 17\%.

To compute the analytic detector response for the three detector geometries, we use \texttt{ASTERIA} \citep{ASTERIA}, a low-energy detector response simulation suite for IceCube that utilises \texttt{SNEWPY} \citep{SNEWS:2021} to simulate the neutrino flux from various CCSN models.

\subsection{Null and signal hypothesis}
\label{sec:ana_hypo}

For our baseline neutrino model, we use the \SI{27}{\solarmass} progenitor Sukhbold model \citep{Sukhbold:2016ccsn} with an LS220 nuclear equation of state (EOS) \citep{Lattimer:1991eos}. This model assumes a progenitor with solar metallicity, with a one-dimensional neutrino transport simulation calibrated to reproduce the observed energy of SN1987A. This model was used as a baseline because it does not exhibit fast-time features with $\mathcal{O}$(\SI{10}{\milli \s}) periodicity and because the integrated neutrino flux is conservative compared to models with similar progenitor masses (see Sec.~\ref{sec:res_syst}). Under the null hypothesis $H_0$, we define the detector hits by the time series
\begin{equation}
    \mathcal{X}_{H_0}(t; d_{\text{SN}}) = N_{\text{SN}}(t; d_{\text{SN}}) + N_{\text{bkg}} \ ,
\end{equation}
which is the sum of the time- and distance-dependent counts from the baseline Sukhbold model and the counts due to the constant detector background.

We add a periodic modification template to the detection hit rate for the fast-time models. The sinusoidal template function, defined between start time $t_0$ and end time $t_1$ with frequency $f$ and amplitude $A$, exhibits integer-multiple periodicity -- i.e., its periodicity aligns with complete sinusoidal cycles. We denote the duration of the periodic modulation as $\Delta T = t_1 - t_0$. Beginning in Sec.~\ref{sec:res} and continuing through the end of this study, will omit the absolute template amplitude $A$ and instead quote the relative amplitude $A' = A/\text{max}\left(\mathcal{X}_{H_0} \right)$ as a percentage of the maximum of $\mathcal{X}_{H_0}$.

Finally, we apply a Hann window $w_{\text{hann}}$ of length $\Delta T$ to the template to smooth out any sharp artefacts:
\begin{equation}
\label{eq:generic}
    N_{\text{ft}}(t) = w_{\text{hann}}(t, \Delta T) \cdot \left[ A \sin(2\pi f t) \cdot H(t_1-t) H(t-t_0) \right] \ ,
\end{equation}
where $H(t)$ is the Heaviside function. The template is added to $\mathcal{X}_{H_0}$, the detector counts under the null hypothesis, to produce the counts under the signal hypothesis $H_1$:
\begin{equation}
    \mathcal{X}_{H_1}(t; d_{\text{SN}}) = N_{\text{SN}}(t; d_{\text{SN}}) + N_{\text{bkg}} + N_{\text{ft}}(t) \ .
\end{equation}

\begin{figure*}
    \centering
    \includegraphics[width=0.7\linewidth]{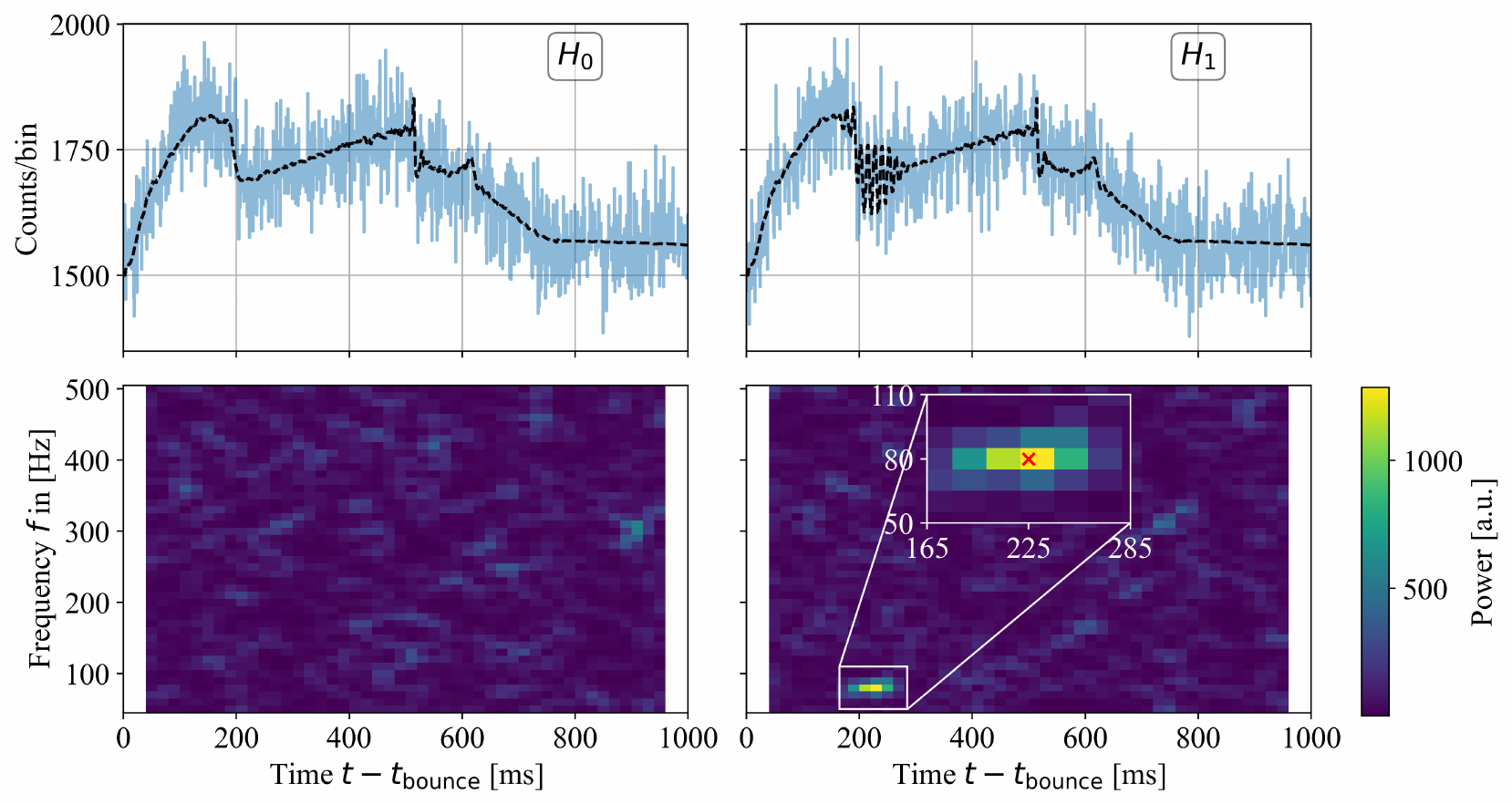}
    \caption{\textit{Top}: simulated signal traces for the baseline Sukhbold model (null hypothesis $H_0$, \textit{left}) and the baseline fast-time model (signal hypothesis $H_1$, \textit{right}) in IceCube for a progenitor \SI{15}{\kilo \parsec} from Earth. Random realisations of the combined detector counts for $H_0$ and $H_1$ are illustrated in blue, while the black, dashed lines indicate the expected counts under $H_0$ and $H_1$. \textit{Bottom}: corresponding Short Time Fourier Transform for $H_0$ (\textit{left}) and $H_1$ (\textit{right}). The red cross in the lower right panel indicates the location of the injected frequency and central time.}
    \label{fig:hypo}
\end{figure*}

To generate random data sets that realise $H_0$ and $H_1$, we randomly draw from Poisson distributions using $\mathcal{X}_{H_0}$ and $\mathcal{X}_{H_1}$ as the respective expectation values.
Since fast-time features occur during the accretion phase of a CCSN, we do not consider signal emission past \SI{1}{\s} post-bounce time. We choose a time resolution of \SI{1}{\milli \s} to resolve frequencies up to the Nyquist frequency of \SI{500}{\Hz}. 
%Figure~\ref{fig:hypo} (upper panel) shows random realisations of the combined detector counts for $H_0$ (left) and $H_1$ (right) in blue, while the black lines indicate the expected counts under $H_0$ and $H_1$.
%The signal traces without the background fluctuation but with an added averaged background, are plotted as a black dashed line on top.
We demonstrate our methodology in Fig.~\ref{fig:hypo}, using a baseline model with $f$=\SI{80}{\Hz} fast-time modulation of 20\% amplitude defined between \SIrange{150}{300}{\milli \s}.

\subsection{Fourier transform and power spectrum}
\label{sec:ana_four}

To extract the power spectrum, we apply a Short Time Fourier Transform (STFT) to our signal traces $\mathcal{X}_{H_0}$ and $\mathcal{X}_{H_1}$, following the methodology in \citet{Walk:2018snft}. By computing the Fourier transform in a time window of width $\Delta t_{\text{wind}}=\SI{100}{\milli \s}$, that slides through the \SI{1000}{\milli \s} long signal in strides of $\Delta t_{\text{slide}}=\SI{20}{\milli \s}$, we obtain the spectral power binned in frequency and time with a resolution of $\delta f=\SI{10}{\Hz}$ and $\delta t=\SI{20}{\milli \s}$, respectively. The width of $\Delta t_{\text{wind}}$ is optimised for the detection of fast-time features. We denote the STFT as $\widetilde{\mathcal{X}}_{H_0}(t,f;d_{\text{SN}})$ and $\widetilde{\mathcal{X}}_{H_1}(t,f;d_{\text{SN}})$ for the null and signal hypothesis, respectively. We apply a Hann window to the sliding window and abstain from zero-padding the signal to avoid edge discontinuities in the Fourier transform. The latter measure crops the signal length by $\Delta t_{\text{wind}}/2=\SI{50}{\ms}$ on both ends of the signal. Finally, we exclude frequencies lower than $f_{\text{cut}}=\SI{50}{\Hz}$ to mitigate the influence of low-frequency components stemming from the overall shape of the light curve. Figure~\ref{fig:hypo} (lower panel) shows the STFT of the signal traces of Fig.~\ref{fig:hypo} (upper panel). The colour scale represents the spectral power proportional to $|\widetilde{\mathcal{X}}|^2$.

\subsection{TS distribution and background trials}
\label{sec:ana_ts}

\begin{figure}
    \centering
    \includegraphics[width=\linewidth]{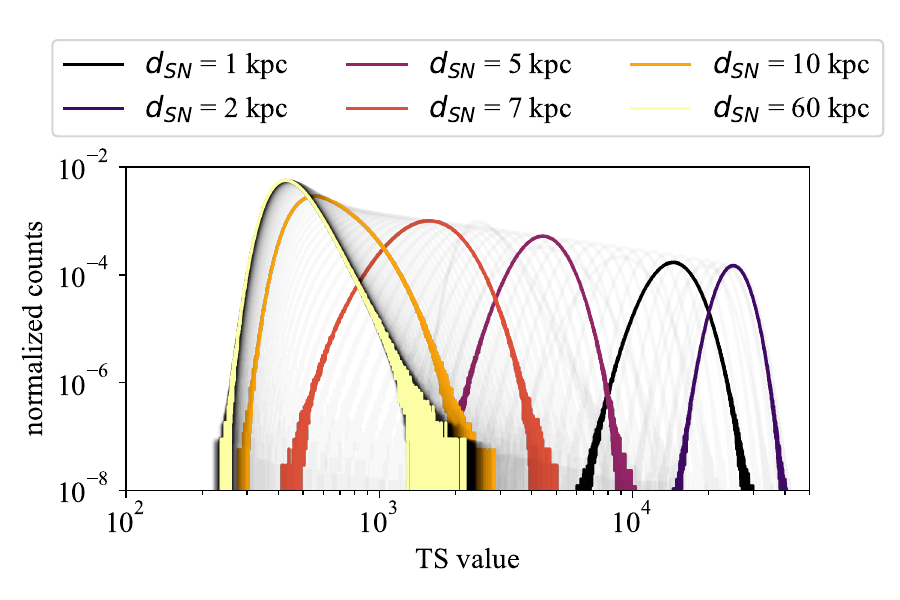}
    \caption{Distance dependence of the binned null hypothesis TS distribution for IceCube for source distances $d_{\text{SN}}$ between \SIrange{0.2}{60}{\kilo \parsec}. The TS distributions reach maximal values around \SI{2}{\kilo \parsec} and shift towards lower TS values for more distant sources.}
    \label{fig:bkg_ts}
\end{figure}

To estimate the sensitivity to fast time variations, we define a test statistic (TS) as the maximum power of the STFT. We perform Monte Carlo pseudo experiments for $H_0$ and $H_1$, generating one TS value per trial. We can then directly infer the significance of the separation between the distributions $\text{TS}_{H_0}=p(\text{TS}|H_0)$ and $\text{TS}_{H_1}=p(\text{TS}|H_1)$. Because $\text{TS}_{H_0}$ has non-Gaussian tails (see Fig.~\ref{fig:bkg_ts}), we use $10^8$ random realisations of $\text{TS}_{H_0}$, compared to $10^4$ realisations of $\text{TS}_{H_1}$, to infer a deviation from $H_0$ of up to $5\sigma$. The p-value for distinguishing $\text{TS}_{H_1}$ from $\text{TS}_{H_0}$ is computed by integrating $\text{TS}_{H_0}$ from quantile $q$, corresponding to a confidence level (CL) of $\text{TS}_{H_1}$, to infinity. As CL we use 16\%, 50\% and 84\%, i.e. the median and the $1\sigma$ Gaussian-equivalent of $H_1$.

Since $H_1$ will always yield greater or equal TS values than $H_0$, we use a one-side Gaussian test to transform the p-value into a Z-score $\xi$. It should be noted that $\xi$ is the significance of discriminating between a model exhibiting fast-time features from one without; it is not the significance of detecting a CCSN. Because $H_0$ is independent of any of the parameters of the generic model, we pre-generate high-statistics trials for $H_0$ between \SIrange{0.2}{60}{\kilo \parsec} in steps of \SI{0.2}{\kilo \parsec} (see Fig.~\ref{fig:bkg_ts}). At close distances, the TS distribution is approximately a log-normal distribution. Peak TS values are reached at a source distance of \SI{2}{\kilo \parsec}, with the distributions shifting to lower TS values and developing non-Gaussian tails for more distant sources. The source distance is binned into steps of \SI{0.2}{\kilo \parsec} to have sufficient resolution to reconstruct the detection horizon for a given CL.

\subsection{Detection significance and detection horizon}
\label{sec:ana_sig}

\begin{figure}
    \centering
    \includegraphics[width=\linewidth]{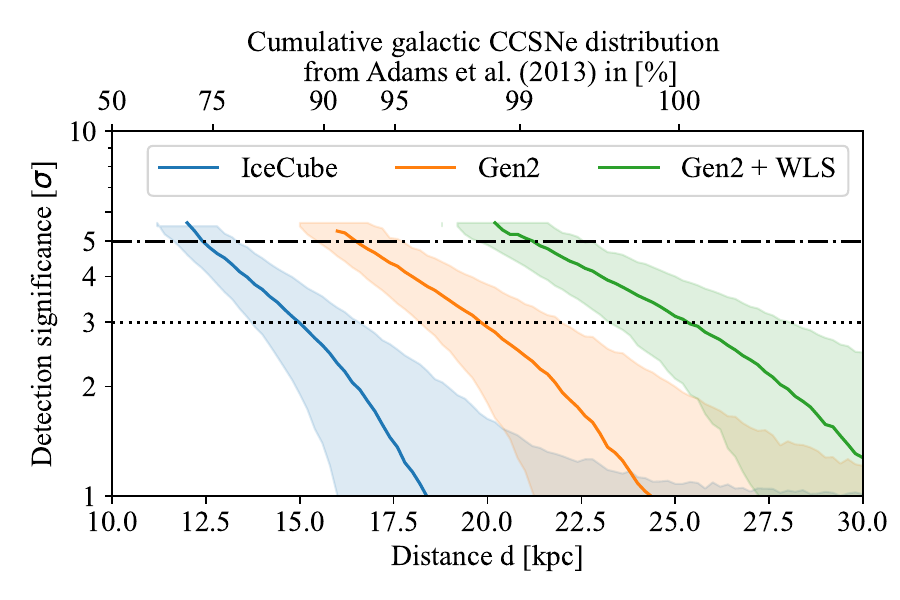}
    \caption{Detection significance versus distance for IceCube (blue), Gen2 (orange) and Gen2+WLS (green) for our baseline fast-time model ($f$=\SI{80}{\Hz}, $A$=20\%). The solid line indicates the 50\% CL of the median significance $\xi$, while the band represents the 16\% and 84\% CL. Indicated here are the $3\sigma$ (dotted line) and $5\sigma$ (dashed line) detection thresholds. The cumulative galactic distribution of CCSNe, excluding the LMC and SMC \citep{Adams:2013rate}, is plotted on the upper horizontal axis. This is related to the detection significance under the assumption that all CCSN models have the same properties as our baseline model.}
    \label{fig:sign_hor}
\end{figure}

The detection significance $\xi(d_{\text{SN}})$ is distance-dependent, and at large distances, the periodic modulation of the neutrino lightcurve becomes indistinguishable from statistical fluctuations in the detector counts. We thus quantify the detection horizon, defined as the distance above which the significance falls below an evidence (discovery) threshold of $3\sigma$ ($5\sigma$) for a given model and detector. Figure~\ref{fig:sign_hor} shows $\xi(d_\text{SN})$ versus distance for the three detector geometries and the baseline fast-time model. The solid lines correspond to the 50\% CLs, while the bands indicate the 16\% and 84\% CLs. For IceCube, the detection significance falls from $5\sigma$ to $3\sigma$ between \SI{12}{\kilo \parsec} and \SI{15}{\kilo \parsec}. To obtain a precise value for the detection horizon, we interpolate the detection significance on the \SI{0.2}{\kilo \parsec} distance grid.

\subsection{Reconstruction of feature parameters}
\label{sec:ana_reco}

The STFT allows us to simultaneously reconstruct the frequency $f_{\text{reco}}$ and temporal position $t_{\text{reco}}$ of the excess power with respect to the injected, true frequency, $f_{\text{true}}$ and central time $t_{\text{true}} = (t_0 + t_1)/2$. We inject $N_{\text{trial}}$ models of \SI{150}{\milli \s} duration with frequency and central time pulled from a uniform distribution defined between $f_{\text{true}} \in [50,~500]~\text{Hz}$ and $t_{\text{true}} \in [125,~875]~\text{Hz}$, respectively. The frequency range is set to account for the low-frequency cut-off at \SI{50}{\Hz}, while the time range is constrained by a combination of the duration of the fast-time feature $\Delta T$ and the size of the window function of $\Delta t_{\text{wind}}/2$. Through bootstrapping, we find that we need $N_{\text{trial}} = 100,000$ to reduce the statistical uncertainty on the most likely values of the reconstructed parameter distribution $(f_\text{reco}, t_\text{reco})$ to below \SI{0.1}{\Hz} and \SI{0.1}{\milli \s} in frequency and central time, respectively.

The resolution of the reconstructed $f_\text{reco}$ and $t_\text{reco}$ is a convolution of the inherent binning of the STFT and the distance-dependent capability of identifying the excess in the power spectrum above statistical fluctuations in the simulated counts. We define the resolution as the $1\sigma$ width (16\% to 84\% quantile) of the marginal distribution $f_{\text{true}}-f_{\text{reco}}$ and $t_{\text{true}}-t_{\text{reco}}$ at a given distance for all geometries.

%__________________________________________________________________

\section{Results}
\label{sec:res}

%We present $\xi(d_\text{SN})$ for a range of fast-time amplitude and frequencies in Sec.~\ref{sec:res_scan}, report the resolution of reconstructing the parameters of the fast-time feature in Sec.~\ref{sec:res_para}, and discuss the effect of model uncertainties in Sec.~\ref{sec:res_syst}.

\subsection{Detection significance}
\label{sec:res_scan}

We consider a fast-time feature of $\Delta T=\SI{150}{\milli \s}$ duration defined between $t_0=\SI{150}{\milli \s}$ and $t_1=\SI{300}{\milli \s}$ with frequencies $f \in[60,~400]~\text{Hz}$ in steps of \SI{10}{\Hz} and amplitudes $A' \in[2.5,~10]\%$ in steps of 2.5\% and $A' \in[10,~40]\%$ in steps of 5\%. In total, we compute the detection horizon for 420 distinct generic fast-time models. For neutrino mixing, we consider the most optimistic case of no flavour transformations during propagation at the source and no Earth matter effects. These settings are revisited in Sec.~\ref{sec:res_syst}. We run 10,000 trials for each generic model and infer $\xi(d_\text{SN})$, as described in Sec.~\ref{sec:ana_sig}.

\begin{figure}
    \centering
    \includegraphics[width=0.9\linewidth]{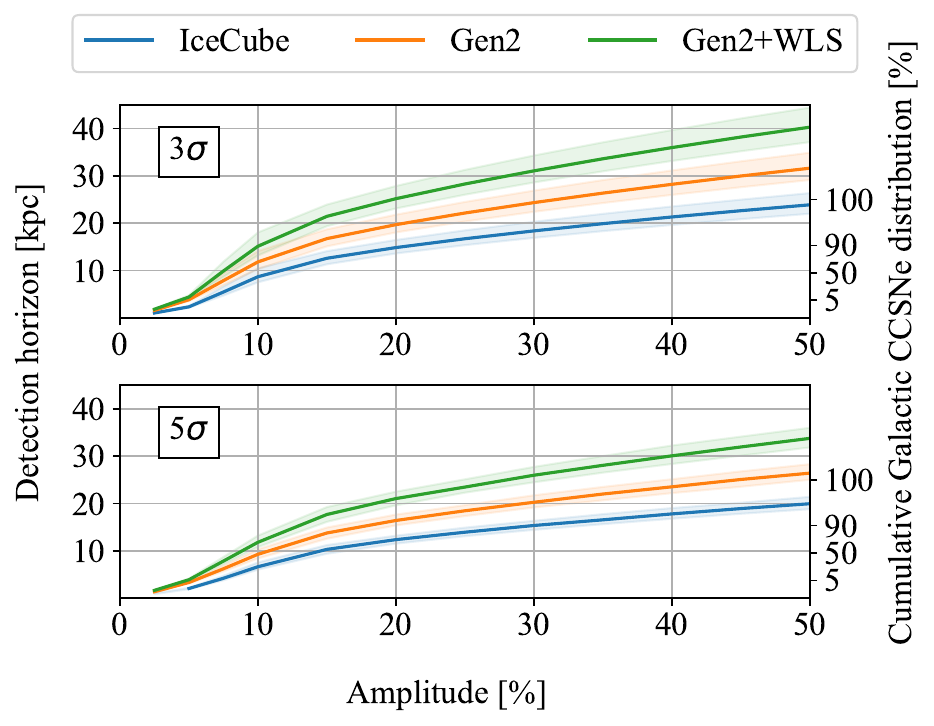}
    \caption{Detection horizon as a function of the relative amplitude for IceCube (blue), Gen2 (orange) and Gen2+WLS (green). \textit{Top}: $3\sigma$ detection horizon. \textit{Bottom}: $5\sigma$ detection horizon. The error band represents the 16\% and 84\% quantiles from Fig.~\ref{fig:sign_hor}. The right vertical axis shows the galactic CCSNe distribution from \citet{Adams:2013rate}.}
    \label{fig:sign_result}
\end{figure}

We find no frequency dependence for the detection horizon of our fast-time features and find an average spread in the frequency-averaged detection horizon of about \SI{0.05}{\kilo \parsec}, \SI{0.07}{\kilo \parsec} and \SI{0.1}{\kilo \parsec} for IceCube, Gen2 and Gen2+WLS respectively. Figure~\ref{fig:sign_result} upper (lower) panel shows the $3\sigma$ ($5\sigma$) detection horizon for the three detector geometries versus the fast-time feature amplitude. The spread from frequency-averaging the detection horizon is negligible. For amplitudes above 15\%, the detection horizon shows a linear behaviour with amplitude. 

\subsection{Reconstruction of feature parameters}
\label{sec:res_para}

We compute the resolution of the reconstructed fast-time feature at the $3\sigma$ and $5\sigma$ detection horizons for a given frequency, amplitude, and detector. The resolution is independent of the injected frequency and central time; therefore, we quote the resolution in absolute rather than relative values. Since the detection horizon for a given amplitude is independent of the injected frequency, we use the frequency-averaged detection horizon from Fig.~\ref{fig:sign_result}. Our simulations demonstrate that the reconstruction resolution is independent of both the detector and the generic model amplitude. This result is expected, as the determining factor is the significance of the excess rather than the detection horizon. The central 68\% distribution of $f_\text{reco}$ at the $3\sigma$ ($5\sigma$) detection level is \SI{7.2}{\Hz} (\SI{7.0}{\Hz}). Similarly, the resolution of the central time $t_\text{reco}$ at the $3\sigma$ ($5\sigma$) threshold is \SI{22}{\milli \s} (\SI{17}{\milli \s}). The uncertainty on the resolution reflects the statistical uncertainty from Fig.~\ref{fig:hypo}.

\subsection{Model uncertainties}
\label{sec:res_syst}

Here we discuss uncertainties arising from the choice of CCSN model, the neutrino flavour conversion mechanism, and detector-related modelling uncertainties. We first discuss their effect on the detection of fast-time features. For some model uncertainties, we only review the impact on the baseline model but do not consider how that affects the detection of fast-time features.

Fast-time features of long duration have more excess power than models with shorter periods. To study this effect, we consider a shorter duration of the fast-time feature, $\Delta T=\SI{50}{\milli \s}$, defined from $t_0=\SI{150}{\milli \s}$ to $t_1=\SI{200}{\milli \s}$ and a longer $\Delta T=\SI{250}{\milli \s}$ duration feature, defined from $t_0=\SI{150}{\milli \s}$ to $t_1=\SI{400}{\milli \s}$.

Neutrino flavour oscillation occurs at the production site at the CCSN and plays an important role in the flavour composition at Earth. Matter effects in dense environments impact the oscillation behaviour and these are particularly relevant for the propagation through the dense core of the CCSN. So far we have neglected neutrino flavour conversion by assuming that the flux composition at Earth is the same as at the production site (no transformation). This represents the most optimistic scenario, as it results in the highest flux of electron antineutrinos that participate in inverse beta decay. To account for the effect of flavour transformations, we consider the conservative scenario where the electron flavour is swapped with the heavy lepton flavour at source (complete exchange), reducing the total detector hit rate by 23\%. Additionally, we consider resonant Mikheev-Smirnov-Wolfenstein (MSW) flavour conversion at the source under adiabatic density conditions, which leads to a reduction in the detection rate of 7\% assuming the normal mass hierarchy (NH) and 23\% for the inverted hierarchy (IH). Earth matter effects are neglected in this work, as their impact strongly depends on the Earth density model and the angle of incidence. However, \citet{IceCube:2024ccsn} reported that MSW oscillation inside Earth can reduce the total detection rate by up to 5\%. 

We report the impact of systematic uncertainties on the detection horizon for a 20\% amplitude model in Tab.~\ref{tab:res_syst}. The duration of the fast-time feature significantly affects the detection horizon. A fast-time feature of \SI{50}{\milli \s} duration can only be seen to half the distance as a fast-time feature of \SI{150}{\milli \s}, while longer features of \SI{250}{\milli \s} duration only slightly increase the detection horizon. Flavour oscillation also influences the detection horizon, with complete flavour exchange resulting in a 16\% reduction of the detection horizon across detectors. Adabatic MSW effects reduce the detection horizon by 9\% (NH) and 16\% (IH). 

To compare the effect that other baseline models would have on the detection of fast-time features, we simulate the number of detected hits in a [100,~400]~ms time window for several models with progenitor masses and equation of state similar to the baseline \SI{27}{\solarmass} Sukhbold model \citep{Sukhbold:2016ccsn}. In particular, we investigate the \SI{27}{\solarmass} Tamborra model \citep{Tamborra:2014sasi} along the direction of maximum SASI, the \SI{27}{\solarmass} O'Connor model \citep{OConnor:2013ccsn}, the \SI{27}{\solarmass} Bollig model with LS220 EOS \citep{Bollig:2015ccsn}, the \SI{27}{\solarmass} Fornax model \citep{Fornax:2020ccsn}, the \SI{27}{\solarmass} Warren model with a turbulent mixing parameter of 1.23 \citep{Warren:2019ccsn}, and the \SI{26}{\solarmass} Zha model \citep{Zha:2021ccsn}. We find that most models yield higher detection rates during the \SI{400}{\milli \s} interval compared to our baseline model: Tamborra shows a 24\% increase, O'Connor a 45\% increase, Fornax a 64\% increase and Zha a 72\% increase. Only the Bolling and Warren model resulted in lower detection rates, with 2\% and 10\% reductions, respectively. Therefore, the results we obtain with the Sukhbold model as our baseline can be considered moderate to conservative in terms of the number of detected events when compared to a range of models with similar progenitor masses.

The dominant uncertainties in the detector simulation arise from uncertainties in the effective volume of the sensors and the interaction cross sections and track lengths of the secondary particles produced in the detector. \citet{IceCube:2011pros} estimates the uncertainty of the photon effective volume on the total detection rate to 12\%—uncertainties related to the dominant inverse-beta decay cross-section account for less than 1\% detection rate. The uncertainty on the mean track length of electrons and positrons in ice has a 5\% effect on the detection rate. Given that the uncertainties on the baseline model are about a factor of 2 to 6 larger, we neglect detector systematic uncertainties.

\begin{table}
    \caption{\label{tab:res_syst}$5\sigma$ detection horizon for a 20\% amplitude model considering various modelling uncertainties. The default model has a $\Delta T = \SI{150}{\ms}$ fast-time feature duration and assumes no flavour transformation.}
    \centering
    \def\arraystretch{1.25}
    \begin{tabular}{lccc}
        \hline\hline
        Model & \multicolumn{3}{c}{$5\sigma$ Detection horizon [kpc]} \\
        Uncertainty & IceCube & Gen2 & Gen2+WLS \\
        \hline
        Default & $12.4_{-0.8}^{+0.8}$ & $16.4_{-1.0}^{+1.2}$ & $21.0_{-1.4}^{+1.5}$ \\
        $\Delta T$ = \SI{50}{\milli \s} & $6.8_{-0.7}^{+0.9}$ & $9.4_{-0.9}^{+1.2}$ & $12.0_{-1.1}^{+1.5}$ \\
        $\Delta T$ = \SI{250}{\milli \s} & $13.1_{-0.7}^{+0.9}$ & $17.4_{-1.0}^{+1.2}$ & $22.4_{-1.3}^{+1.5}$ \\
        Complete Exchange & $10.4_{-0.6}^{+0.8}$ & $13.7_{-0.8}^{+1.0}$ & $17.6_{-1.0}^{+1.3}$ \\
        Adiabatic MSW NH & $11.4_{-0.7}^{+0.8}$ & $15.0_{-0.9}^{+1.1}$ & $19.3_{-1.1}^{+1.4}$ \\
        Adiabatic MSW IH & $10.4_{-0.6}^{+0.8}$ & $13.7_{-0.8}^{+1.0}$ & $17.7_{-1.0}^{+1.2}$ \\
        \hline
    \end{tabular}
    \def\arraystretch{1.0}
\end{table}

%______________________________________________________________

\section{Discussion}
\label{sec:disc}

For IceCube-Gen2 equipped with wavelength shifters (Gen2+WLS), we will detect faint fast-time features in the entire Milky Way at $>3\sigma$ ($>5\sigma$) for models with amplitudes down to $A \geq  20\%$ ($A \geq  25\%$). This is a significant improvement over the current capability of IceCube, where we can only detect modulations of $A \geq  50\%$ at $3\sigma$. IceCube-Gen2 with wavelength shifters also offers a factor of 2 improvement over the baseline design of IceCube-Gen2, which is sensitive to amplitudes of $A \geq  30\%$ ($A \geq  40\%$). 

If a fast-time feature is detected at the $3\sigma$ ($5\sigma$) level, we find that its frequency and central time can be reconstructed independently of the frequency, amplitude and detector geometry with \SI{7.2}{\Hz} (\SI{7.0}{\Hz}) and \SI{22}{\milli \s} (\SI{17}{\milli \s}), respectively.

We compare our results to the detection horizon for fast-time features reported in \citet{Beise:2023sasi}, which used a time-integrated Fourier analysis. This earlier study identified a $5 \sigma$ detection horizon (excluding systematics) for the \SI{27}{\solarmass} (\SI{20}{\solarmass}) Tamborra model \citep{Tamborra:2014sasi}, approximately corresponding to a 10\%-15\% (20\%) generic amplitude model, as \SI{9}{\kilo \parsec} (\SI{13}{\kilo \parsec}) for IceCube and \SI{13}{\kilo \parsec} (\SI{20}{\kilo \parsec}) for IceCube-Gen2+WLS. This sensitivity is in good agreement with the results presented in this paper. 

KM3NeT reports a $3\sigma$ detection horizon for fast-time features of \SI{3}{\kilo \parsec} for the \SI{27}{\solarmass} model for the combined ARCA and ORCA detector \citep{Molla:2019snft}, which is roughly half of IceCube's sensitivity of about \SI{11}{\kilo \parsec} for the same model \citep{Beise:2023lice}.

In this study, we probed for mono-frequency fast-time features with constant amplitude. More complex features with several potentially time-dependent frequency modes would smear the excess power across several bins in the Fourier spectrum and would thereby reduce the sensitivity. By applying a threshold frequency of $>\SI{50}{\Hz}$, our analysis excludes low-frequency fast-time features, which in any case are not predicted by existing models. As mentioned in Sec.~\ref{sec:res_syst}, IceCube's sensitivity depends on the baseline model, which is not known a priori. However, we show that the selected baseline model predicts a moderate number of events in the detector compared to other models with similar progenitor masses.

There is potential for further improvement by fully utilising the capabilities of coincidence triggers in segmented sensors. While we have not simulated coincidence triggers in this work, \citet{LozanoMariscal:2021mult} showed that coincidence triggers in the mDOM can drastically reduce uncorrelated and correlated background counts while still preserving most of the CCSN signal. 

%______________________________________________________________

\section{Conclusions}
\label{sec:conc}

Measuring the neutrino flux of the next galactic CCSN with high statistics will reveal crucial information about the fundamental hydrodynamical processes involved in the explosion mechanism of the CCSN, which can manifest as fast-time features in the observed neutrino flux. Fast-time features in the flux are predicted to arise from hydrodynamical instabilities during the accretion phase or fast rotations of the progenitor, with model predictions varying greatly. IceCube and the planned large-scale extension IceCube-Gen2 would cover a large parameter space of generic fast-time features, with IceCube-Gen2 having greater sensitivity to fainter modulations and Galactic CCSNe at larger distances.

Wavelength-shifting technology has the potential to improve the sensitivity while constituting a comparatively inexpensive addition to the prospective sensor deployment. IceCube-Gen2, especially if equipped with wavelength shifters, will significantly enhance the detection of fast-time features throughout the Milky Way, allowing for $>5\sigma$ detection of modulations as low as 25\% of the maximum flux, with the frequency and central time of these features reconstructable to within \SI{7}{\Hz} and \SI{17}{\milli \s}, respectively. The IceCube-Gen2 values demonstrate significant improvements to the detector's sensitivity and reconstruction capabilities compared to IceCube.

\begin{acknowledgements}
We acknowledge fruitful discussions with the Swedish IceCube group, and in particular with Alan Coleman.
The computations and data handling were enabled by resources provided by the National Academic Infrastructure for Supercomputing in Sweden (NAISS), partially funded by the Swedish Research Council through grant agreement no. 2022-06725.

\end{acknowledgements}


% for the bibliography, at the end
\bibliographystyle{aa} % style aa.bst
\bibliography{references.bib} % your references 

\end{document}